\documentclass[graybox]{svmult}

\usepackage{mathptmx}       % selects Times Roman as basic font
\usepackage{helvet}         % selects Helvetica as sans-serif font
\usepackage{courier}        % selects Courier as typewriter font
\usepackage{type1cm}        % activate if the above 3 fonts are
\usepackage{makeidx}         % allows index generation
\usepackage{graphicx}        % standard LaTeX graphics tool
\usepackage{multicol}        % used for the two-column index
\usepackage[bottom]{footmisc}% places footnotes at page bottom

\makeindex             % used for the subject index
\begin{document}

\title*{The WMAP cold spot}
\author{M. Cruz, E. Mart\'{\i}nez-Gonz\'alez, P. Vielva}
\institute{M.Cruz \at IFCA, CSIC-Univ. de Cantabria, Avda. los Castros, s/n, 39005-Santander, Spain, \email{cruz@ifca.unican.es}
\and E. Mart\'{\i}nez-Gonz\'alez \at IFCA, CSIC-Univ. de Cantabria, Avda. los Castros, s/n, 39005-Santander, Spain, \email{martinez@ifca.unican.es}
\and P. Vielva \at IFCA, CSIC-Univ. de Cantabria, Avda. los Castros, s/n, 39005-Santander, Spain, \email{vielva@ifca.unican.es}}
\maketitle

\abstract*{The WMAP cold spot was found by applying spherical wavelets to the first year WMAP data. An excess of kurtosis of the wavelet coefficient was observed at angular scales of around 5 degrees. This excess was shown to be inconsistent with Gaussian simulations with a p-value of around $1\%$. A cold spot centered at ($b = -57^\circ, l = 209^\circ$) was shown to be the main cause of this deviation. Several hypotheses were raised to explain the origin of the cold spot. After performing a Bayesian template fit a collapsing cosmic texture was found to be the most probable hypothesis explaining the spot. Here we review the properties of the cold spot and the possible explanations.}

\abstract{The WMAP cold spot was found by applying spherical wavelets to the first year WMAP data. An excess of kurtosis of the wavelet coefficient was observed at angular scales of around 5 degrees. This excess was shown to be inconsistent with Gaussian simulations with a p-value of around $1\%$. A cold spot centered at ($b = -57^\circ, l = 209^\circ$) was shown to be the main cause of this deviation. Several hypotheses were raised to explain the origin of the cold spot. After performing a Bayesian template fit a collapsing cosmic texture was found to be the most probable hypothesis explaining the spot. Here we review the properties of the cold spot and the possible explanations.}

\section{Introduction}
\label{sec:1}
The Gaussianity of the Cosmic Microwave Background (CMB) anisotropies is a testable prediction of the simplest inflationary models. %(Bardeen et al. 1983, Guth \& Pi 1985). 
Alternative models such as non-standard inflation or topological defects predict non-Gaussian features. Therefore a
Gaussianity analysis of the temperature anisotropies in the CMB can help us to discriminate between different cosmological models. 
The Wilkinson Microwave Anisotropy Probe mission (WMAP, Bennett et al. 2003), measured the CMB fluctuations with high accuracy. 
In a first approach (Komatsu et al. 2003), these were found to be consistent with the Gaussian predictions. 
Further analyses revealed asymmetries or non-Gaussian features which have been confirmed in later releases of the WMAP data (Hinshaw et al. 2007, Hinshaw et al. 2008). 
Some of these anomalies are: low multipole alignments (de Oliveira-Costa et al. 2006, Copi et al. 2007, Land \& Magueijo 2007, Chiang et al. 2007), 
north-south asymmetries (Eriksen et al. 2007, Bernui et al. 2007),  structure alignment (Vielva et al. 2007), low variance (Monteser\'{\i}n et al. 2008 ), non-Gaussian features found with steerable (Wiaux et al. 2008) 
and directional wavelets (McEwen et al. 2006), and a very cold spot in the southern hemisphere (Cruz et al. 2007a). 
This cold spot (CS) is found at Galactic coordinates ($b = -57^\circ, l = 209^\circ$), and has an angular radius in the sky of about $5^\circ$. 

\section{Detection and significance}
\label{sec:2}
The CS was found by applying the Spherical Mexican Hat Wavelet (SMHW, Mart\'{\i}nez-Gonz\'alez et al. 2002) to the WMAP data (Bennett et al. 2003).
Convolving a CMB map with the SMHW at a particular wavelet scale increases the signal to noise ratio at that scale. Moreover, 
the spatial location of the different features of a map is preserved. It is an optimal tool to enhance circular non-Gaussian features on the sphere. 

Vielva et al. (2004) performed a blind Gaussianity analysis, applying the SMHW to the 1-year WMAP data. 
The analysis was performed on 15 arbitrary wavelet scales and two estimators, skewness and kurtosis, were used. 
The kurtosis deviated significantly from the range of values expected from Gaussian simulations at scales around $5^\circ$. The most
outstanding feature at that scale was the CS. The spot was found to be the main cause of the deviation from Gaussianity, using the area estimator (Cruz et al. 2005).

The morphology of the CS was studied applying wavelets with different ellipticities (Cruz et al. 2006). The shape of the CS was found to be
almost circular. A multifrequency analysis revealed that the CS is frequency independent (Cruz et al. 2006).

Cay\'on, Jin \& Treaster (2005) redetected the spot combining wavelet analyses with two new estimators, namely $Max$ and Higher Criticism.
Cruz et al. (2007a) confirmed the detection in the 3-year WMAP data, using all the mentioned estimators and performed a rigorous significance analysis. 
The probability of finding an at least as high deviation in Gaussian simulations was calculated without any
\emph{a posteriori} assumption. Thus, only the blind analysis using the skewness and kurtosis estimators was taken into account, since further tests can be regarded
as follow-up tests. A conservative estimate of this probability was $1.85\%$ considering the skweness and kurtosis estimators. 
Considering only the kurtosis estimator is also a valid approach and gives $0.86\%$. Hence the CS shows a significant deviation from Gaussianity.

Further wavelet analyses redetected the CS using steerable wavelets (Vielva et al. 2007) or needlets (Pietrobon et al. 2008). R{\"a}th et al.(2007) detected the CS in the 3-year WMAP data using scalar indices instead of wavelets.

\section{Origin}
\label{sec:3}
The first analyses on the origin of the CS were focused on checking the contribution of systematic effects or foregrounds to the total detected signal.
Vielva et al. (2004) and Cruz et al. (2005) showed that all the detectors measured an almost constant amplitude of the CS and proved that the instrumental
noise was not affecting the results. Mukherjee \& Wang (2004) found that the results were robust using different exclusion masks. A careful foreground analysis 
was performed in Cruz et al. (2006). The flat frequency dependence of the CS implied that foreground residuals were very unlikely to explain this outstanding spot.

A number of models were proposed in order to explain this anomaly: voids (Inoue \& Silk 2006, 2007, Rudnick, Brown \& Williams 2007, Granett et al. 2008), 
second order gravitational effects (Tomita 2005, Tomita \& Inoue 2007), anisotropic Bianchi VII$_h$ model (Jaffe et al. 2005), finite cosmology model, (Adler, Bjorken \& Overduin 2006), asymptotically flat Lemaitre-Tolman-Bondi model (Garcia-Bellido \& Haugbolle 2008, Masina \& Notari 2008) and brane-world model (Cembranos et al. 2008) are some of them. 
Also a large-scale, non-Gaussian angular modulation (Naselsky et al. 2007) has been suggested although this hypothesis is not based on any physical model.
The Bianchi VII$_h$ model was proven by Jaffe et al. (2006) to be ruled out at high 
significance, and there is still no further evidence for the validity of the other explanations. A full Bayesian evidence
analysis of that Bianchi model can be found in Bridges et al. (2007a). Bridges et al. (2007b) show that the cold spot was likely to be driving any Bianchi VII$_h$ template detection.

Recently, Cruz et al. (2007b) showed through a Bayesian statistical analysis that the CS could be caused by a cosmic texture (Turok 1989), 
a type of cosmic defect which arises when a simple Lie group like SU(2) is completely broken. 
The amplitude and scale of the spot were consistent with that interpretation and the kurtosis of the data was compatible with
the Gaussian CMB plus texture model at all scales. The predicted number of spots at a scale of $5^\circ$ or larger, is $\approx 1$ which is consistent with the
observed spot. In Figure 1 we show the CS in the WMAP data, the best fit texture template and the data after subtracting the template.

\begin{figure}[t]
\sidecaption[t]
\includegraphics[scale=.35]{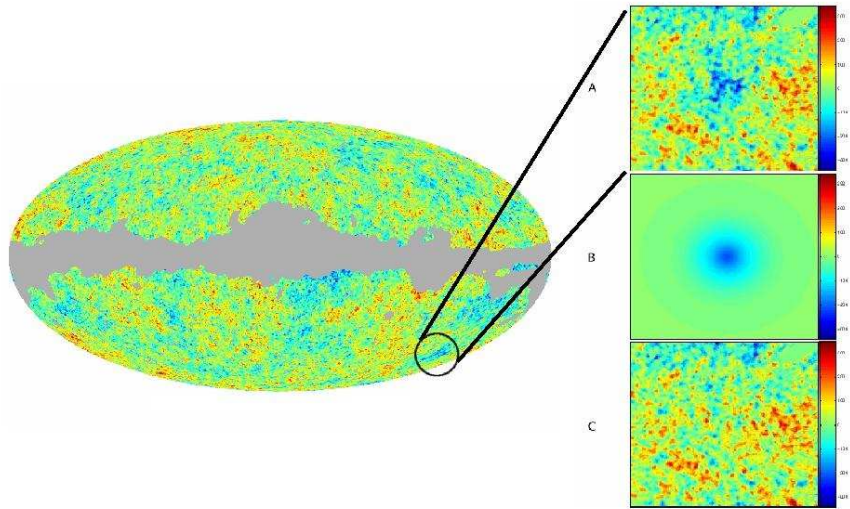}
%
% If no graphics program available, insert a blank space i.e. use
%\picplace{5cm}{2cm} % Give the correct figure height and width in cm
%
\caption{Combined frequency map of the Wilkinson Microwave Anisotropy Probe (WMAP) in Galactic coordinates. 
The region of the sky most contaminated by our own Galaxy is masked (gray pixels). The region of the cold spot is marked with a black circle.
A: Azimuthal projection of a 43 times 43 degrees patch  centered at the spot (Galactic coordinates b= -57, l= 209 degrees). 
B: Best fit template induced by a cosmic texture. 
C: The map in A after subtracting the texture template. The temperature units shown in the colorbars are microKelvin and the pixel is $13.7$ arcmin.}
\label{fig:1}       % Give a unique label
\end{figure}

Cruz et al. (2008) perform a Bayesian template fit on two other possibilities, namely the Rees-Sciama (Rees \& Sciama 1968) effect caused by voids as proposed by Inoue \& Silk (2006, 2007) or Rudnick, Brown \& Williams (2007) and the Sunyaev-Zeldovich effect produced by clusters (hereafter SZ effect, Sunyaev \& Zeldovich 1970) already 
considered in Cruz et al. (2005). The results obtained for these two models are then compared to the cosmic texture model results and the most probable hypothesis is shown to be the cosmic texture one. The void and cluster hypotheses are discarded since they need unrealistic values for their parameters.

\section{Discussion}
\label{sec:4}
Cruz et al. (2007b, 2008) showed recently that the CS could be due to a collapsing texture. 
They performed a Bayesian template fit on the 5-year WMAP data (Hinshaw et al. 2008), using an approximated temperature profile based on the analytic 
result given in Turok \& Spergel (1990). 
The Bayesian posterior probability ratio was used to compare the null hypothesis (which interpreted the CS as Gaussian CMB), to the alternative hypothesis 
(Gaussian CMB plus cosmic texture). The analysis favoured the texture hypothesis. Applying the same analysis to extreme Gaussian CMB spots taken from simulations,
showed that around 5\% of these Gaussian CMB spots imply higher posterior probability ratios.

The scale and amplitude of the spot ($5.1^\circ$ and $dT/T = 7.7 \times 10^{-5}$) were consistent with having being caused by a texture 
at redshift $z \sim 6$ and energy scale $\phi_0 \approx 8.7 \times 10^{15} GeV$. 
The number of expected texture spots of scale $5.1^\circ$ or above is around $1.1$ consistent with the single observed spot.

The amplitude of the texture spot was shown to be overestimated by a selection bias (Cruz et al. 2007b). Texture spots with amplitude $4 \times 10^{-5}$ were simulated 
and added to CMB Gaussian simulations. Performing a template fit on these single extreme events, the inferred amplitude was in average $7.9 \times 10^{-5}$, i.e. 
almost twice the true amplitude.
This means that the amplitude required by the CS is not in conflict with the upper limits on the amplitude calculated from the power 
spectrum (Urrestilla et al. 2007).

Moreover, Bevis et al. (2008) and Urrestilla et al. (2007) find a $2\sigma$ preference for a cosmological model including texture or strings, 
when considering only CMB data. However this preference is reduced when including Hubble Key Project (Freedman et al. 2001) and Big Bang 
nucleosynthesis (Kirkman et al. 2003) constraints in the Bayesian power spectrum analysis.
%%%

The finding of a local supergroup (Brough et al. 2006) and a dip in the NVSS survey (Rudnick, Brown \& Williams 2007), which match the position
of the CS could be a hint for the cluster or void hypotheses, but a detailed Bayesian analysis reveals that these are statistical coincidences 
due to a posteriori analyses (Cruz et al. 2008, Smith \& Huterer 2008).

The large angular size and temperature decrement of the CS are hard to explain with conventional clusters or voids. 
We would need either a local, rich cluster or a huge void with radius $\sim 100-200 h^{-1}$ Mpc located at redshift $z \le 1$.
In the cluster case only a local group with Compton parameter $y_o < 10^{-7}$ is observed in the required direction. 
The angular scale is marginally consistent with the spot but the temperature decrement generated is at most a few $\mu K$ and 
hence cannot account for the CS. Our Bayesian analysis gives $\rho \approx 0.05$, discarding the cluster hypothesis.

The temperature decrements produced by conventional voids with radii around $10 h^{-1}$ Mpc are negligible.
Explaining the cold spot by a void with radius bigger than $100 h^{-1}$ Mpc, would introduce a $13\sigma$ anomaly in the standard model to solve the $\approx 99\%$ anomaly of the spot. The solution would be worse than the problem.

Hence we can conclude that the texture hypothesis is the most plausible explanation for the CS. 

High resolution temperature and polarization measurements will be able to discriminate among the void and texture hypotheses (Das \& Spergel 2008). Both of them produce a specific lensing effect in the CMB. Textures do not have an associated polarization signal while Gaussian CMB spots have a correlated polarization pattern.
In addition the texture model predicts more CMB texture spots at scales around $1-2^\circ$. These are more difficult to detect because the Gaussian CMB reaches its maximum power at these scales, but it could be possible to detect them through an all-sky Bayesian analysis, which is already in progress.

%%%%%%%%%%%%%%%%%%%%%%%% referenc.tex %%%%%%%%%%%%%%%%%%%%%%%%%%%%%%
% sample references
% %
% Use this file as a template for your own input.
%
%%%%%%%%%%%%%%%%%%%%%%%% Springer-Verlag %%%%%%%%%%%%%%%%%%%%%%%%%%
%
% BibTeX users please use
% \bibliographystyle{}
% \bibliography{}
%

\end{document}